\documentclass[aoas,MSNbibl,nameyear,seceqn,dvips]{arximspdf}
\usepackage{algorithmic}
\usepackage{algorithm}
\usepackage{graphicx}

\doi{10.1214/12-AOAS548} 
\volume{6}
\issue{3}
\pubyear{2012}
\firstpage{1162}
\lastpage{1184}

\makeatletter

\newcommand{\T}{\top}
\newcommand{\ph}{\hat{p}}
\newcommand{\pmin}{p_{\min}}
\newcommand{\Tt}{{\widetilde{T}}}
\newcommand{\argmax}{\mathop{\arg\max}}
\newcommand{\dof}{\operatorname{DF}}
\newcommand{\difftree}{{difftree}}
\newcommand{\pt}{\tilde{p}}

\newcommand{\return}{\STATE\textbf{return}\ }

\makeatother

\begin{document}
\begin{frontmatter}

\title{Tree models for difference and change detection in a complex environment}
\runtitle{Tree models for difference and change detection}

\begin{aug}
\author[A]{\fnms{Yong} \snm{Wang}\corref{}\ead[label=e1]{yongwang@auckland.ac.nz}},
\author[A]{\fnms{Ilze} \snm{Ziedins}\ead[label=e2]{i.ziedins@auckland.ac.nz}},
\author[A]{\fnms{Mark} \snm{Holmes}\ead[label=e3]{m.holmes@auckland.ac.nz}}
\and
\author[B]{\fnms{Neil} \snm{Challands}\ead[label=e4]{Neil.Challands@fire.org.nz}}
\runauthor{Wang, Ziedins, Holmes and Challands}
\affiliation{University of Auckland, University of Auckland,
University of Auckland and~New~Zealand Fire Service}
\address[A]{Y. Wang\\
I. Ziedins\\
M. Holmes\\
Department of Statistics\\
University of Auckland \\
Private Bag 92019, Auckland\\
New Zealand\\
\printead{e1}\\
\hphantom{E-mail: }\printead*{e2}\\
\hphantom{E-mail: }\printead*{e3}} 
\address[B]{N. Challands\\
New Zealand Fire Service\\
P.O. Box 2133, Wellington\\
New Zealand\\
\printead{e4}}
\end{aug}

\received{\smonth{3} \syear{2011}}
\revised{\smonth{12} \syear{2011}}

%
\begin{abstract}
A new family of tree models is proposed, which we call ``differential
trees.'' A differential tree model is constructed from multiple data sets
and aims to detect distributional differences between them. The new
methodology differs from the existing difference and change detection
techniques in its nonparametric nature, model construction from multiple
data sets, and applicability to high-dimensional data. Through a detailed
study of an arson case in New Zealand, where an individual is known to
have been laying vegetation fires within a~certain time period, we
illustrate how these models can help detect changes in the frequencies of
event occurrences and uncover unusual clusters of events in a complex
environment.
\end{abstract}

%

\end{frontmatter}

\section{Introduction}
\label{secintro}

We propose a new family of tree models that can be used to uncover
distributional differences between multiple data sets. These models, which
we call ``differential trees,'' are suitable for solving sophisticated,
multivariate problems. They can be applied, for instance, to change
detection and work effectively in an online surveillance fashion.

The research was motivated by a real-world problem. Fire service
departments are often interested in detecting changes in the
frequencies of
different types of fire incidents, automatically from large amounts of data
and informatively to shed light on potential causes. This change detection
problem is certainly not unique to fire incidents. Similar problems can
easily be found in many fields such as climatology, epidemiology and
economics.

To investigate the problem in depth, one particular scenario has been
chosen as a case study, and is used exclusively in this paper to illustrate
and investigate the new methodology. It was known that an individual had
been laying vegetation fires between October 2006 and January 2007 in the
urban area of Blenheim, New Zealand. The New Zealand Fire Service wishes
to be able to automatically detect such a sequence of events as early as
possible and isolate them from the rest. At first glance, there seems to
be a lack of information to relate the scenario to frequency change
detection, since no fire maliciously set by an individual could be
definitely known as such in reality. However, a surrogate variable can be
used. All fire incidents are categorized by on-the-spot fire fighters as
either suspicious or not. Since the maliciously-set fires should be highly
correlated to those labeled suspicious, we turn the vaguely-defined
practical problem into one of detecting changes in the frequencies of: (a)
suspicious and other fires, (b) suspicious fires only, as a more direct
approach, or (c) fire incidents of a different categorization, as a less
direct approach. We consider the frequency changes as distributional
differences.

The problem above poses a number of challenges for traditional change
detection methods that rely on parametric assumptions
[\citet{basseville-nikiforov1993},
\citet{gustafsson2000}, \citet{poor-hadjiliadis2009}]. For
this and similar problems, there may exist a number of potentially relevant
variables, which can be either numerical or categorical and may contain
missing values. The distribution of fire incidents may depend on many
factors, such as geographical, seasonal, time-of-day and day-of-week
effects, and is simply impossible to model parametrically. Moreover, an
arsonist may operate in certain time periods and in certain neighborhoods,
and light fires of certain types.

By contrast, the proposed methodology is particularly suitable for solving
such problems. Though belonging to the family of tree models
[\citet{breiman-etal1984}, \citet{morgan-sonquist1963},
\citet{quinlan1993}], a differential
tree is constructed from multiple data sets, as opposed to from a single
data set by a conventional method, and purpose-built for difference
detection. Intuitively, the method stacks the data sets on top of one another
(imagine a two-dimensional case) and then, via recursive space partitioning
of tree-structured models, looks for the local areas with
heterogeneity. By
ignoring variations in individual data sets that are common to all and thus
irrelevant to changes, such as geographical and seasonal effects in the
arson case, it makes more efficient use of data information than an
approach that builds one model from each data set. Hence, it achieves a
gain in power which is similar in spirit to that of the paired $t$-test or
blocking in experimental design.

\begin{table}
\caption{Variables}
\label{tabvariables}
\begin{tabular*}{\tablewidth}{@{\extracolsep{\fill}}lp{75pt}p{190pt}@{}}
\hline
\textbf{Name} & \multicolumn{1}{c}{\textbf{Meaning}} & \multicolumn{1}{c@{}}{\textbf{Values}} \\
\hline
\texttt{x} & Map grid east & Real \\
\texttt{y} & Map grid north & Real \\
\texttt{Urban} & \mbox{Whether an urban} \mbox{or rural area}
& $\{1 = \mbox{urban}$, $0 =\mbox{rural}\}$ \\[2pt]
\texttt{Alarm} & \mbox{Alarm method} \mbox{code} &
\mbox{$\{1 = 111$ emergency call, $2 =\mbox{exchange}$
phone call}, \mbox{$3 = \mbox{running}$ call,
$4 = \mbox{police}$/ambulance}, \mbox{$5 = \mbox{private}$ fire alarm,
$6 =\mbox{other}\}$}\\[2pt]
\texttt{Firetype} & \mbox{Type of fire} \mbox{incident}
& $\{1 = \mbox{structure}$, $2 =
\mbox{mobile}$ property, $3 =\mbox{vegetation}$,
$4 = \mbox{chemical}$, $5 = \mbox{rubbish}$, $6 = \mbox{other}\}$\\
[2pt]
\texttt{Heatsource} & Heat source & $\{1 = \mbox{outside}$ fire lit for
lawful purpose,
$2 = \mbox{gas}$/liquid fuelled equipment, $3 = \mbox{solid}$ fuelled equipment,
$4 = \mbox{electrical}$ equipment, $5 = \mbox{hot}$ object,
$6 = \mbox{fireworks}$,
$7 = \mbox{cigarette}$/smoking materials, $8 = \mbox{act}$ of nature,
$9 = \mbox{exposure fire}\}$ \\
[2pt]
\texttt{Objignited} & Object ignited & $\{1 = \mbox{structure}$ component,
$2 =
\mbox{furniture}$/appliances,
$3 = \mbox{soft}$ goods/bedding, $4 = \mbox{decoration}$/recreational
\mbox{materials,
$5 = \mbox{storage}$ containers and
materials,}
\mbox{$6 = \mbox{electrical}$ equipments/tyres/insulators,} \mbox{$7 =
\mbox{outdoor}$ items,
$8 = \mbox{hazardous}$ substances} \mbox{and fuels, $9 = \mbox{other}\}$} \\
[2pt]
\texttt{Time} & \mbox{Time of day} & $[0, 24)$ \\
\texttt{day} & \mbox{Day of the} \mbox{quadrennial} \mbox{period}
& $\{1, 2, \ldots, 1461\}$\\
[2pt]
\texttt{Dayweek} & Day of the week & $\{1 =  \mbox{Monday}, \ldots, 7 =
\mbox{Sunday}\}$\\
\texttt{label} & \mbox{Category labeled} \mbox{by fire fighters}&
$\{$\texttt{suspicious}${}={}$suspicious fire, \texttt{other}${}={}$other
type$\}$ \\
\hline
\end{tabular*}
\end{table}

The arson data used throughout the paper contains information for all fire
incidents that occurred within and around Blenheim, a moderately sized town
(population 30,200), between 1/Jan/2004 and 31/Dec/2007,
as stored in $11$ variables, with names, meanings and possible values
given in
Table~\ref{tabvariables}. During the quadrennial period, there were a
total of $704$ fire incidents, $171$ of which were labeled suspicious. Two
variables, \texttt{heatsource} and \texttt{objignited}, contain, respectively,
$342$ and $275$ missing values. Pairs of disjoint subsets of the data will
be produced in various ways below, and will be used to construct
differential trees. Our main focus will be on contrasting the subsets
in two biennial periods,
2004--2005 and 2006--2007, to uncover the unusual cluster(s) of fire
incidents in the latter period that are likely related to the arson case.
We shall also apply the methodology in a sequential detection fashion and
compare two consecutive annual periods by shifting time periods
progressively. Random subsets will also be produced by permutation or
bootstrapping for assessing and enhancing performance.

The rest of the paper is organized as follows. Section~\ref{secoverview}
briefly reviews the problem of change detection and tree models and gives
an overview of the proposed methodology. Section~\ref{secdifftree}
describes in detail the differential tree models and their
construction. A
primary study of the arson case is presented in
Section~\ref{secarson}. The performance of the method will be assessed and
enhanced in Section~\ref{secperform}, with an application in a sequential
detection fashion given in Section~\ref{secsequential}.
Section~\ref{secfurther-studies} investigates building differential trees
using other responses, and Section~\ref{secsummary} gives some concluding
remarks.

The data and computer code for carrying out the analysis presented in the
paper are given in the supplementary material [\citet{wang-etal2012}].

\section{An overview}
\label{secoverview}

\subsection{Change detection}
\label{secchange-detection}

Change detection has a long history of research in statistics, with a focus
on detecting change points
[\citet{lai1995}, \citet{maceachern-etal2007},
\citet{page1954}, \citet{shewhart1931}].
These methods, however, rely on parametric assumptions and are applied to
situations, such as industrial process control, where such assumptions can
be safely made.

Another very useful technique for detecting changes is scan statistics
[\citet{glaz-etal2001}, \citet{naus1965}]. This technique looks for unusual
clusters of temporal or spatial events in a~single data set by using
a~scanning window to locate clusters of observations that differ in
distribution from the rest. Because of the high computational cost, it is
only applicable to low-dimensional problems.

\subsection{Tree models}
\label{sectree}

Tree models are often used to solve difficult, high-dimen\-sional
problems. There are two major families, classification and regression
trees, for a categorical and a continuous response variable, respectively
[\citet{breiman-etal1984}]. Other families also exist but are less used,
for example, Poisson regression trees for a count response
[\citet{chaudhuri-etal1995}, \citet{therneau-atkinson1997}] and survival trees
for a
failure time response with censoring
[\citet{davis-anderson1989}, \citet{ishwaran-etal2008}].

As in the references above, the basic idea of tree modeling is to
partition the space of explanatory
variables recursively into increasingly smaller regions so that a simple
model fits well to the data in a minimal region. We call such a simple
model an \textit{atomic model}, which can be, for example, the constant
function or
the normal distribution for a continuous response, or a multinomial
distribution for a categorical response, as for regression and
classification trees, respectively. A tree model is the composite of the
atomic models in the minimal regions and is best represented by a rooted
tree, in which a node corresponds to a region, a terminal node a minimal
region, and the branching under an internal node a space partitioning.
Each internal node thus also has a \textit{subtree model}.

Building a tree model typically consists of splitting and pruning
stages. Splitting proceeds in a top-down fashion, by selecting at each node
a split in the form of a logical condition from a large number of
candidates, which aims to maximize\vadjust{\goodbreak} the homogeneity in subregions.
Univariate binary splits are commonly used, for example, $x \le3.5$
for a
continuous variable, or \texttt{season}${}={}$\{\texttt{spring},
\texttt{autumn}\}
for a categorical variable. Splitting continues until homogeneity is
reached in a region. An exhaustive splitting is generally beneficial, and
allows for uncovering relations hidden deep under the surface. However,
a tree
grown only by splitting is likely to overfit the data. Hence, it is often
followed by pruning, which replaces spurious subtrees with their root nodes
in a bottom-up fashion.

Terminal nodes are important for a tree model and the features of
interest at those nodes are described by the atomic models. We shall
often use the word
``pattern'' to specifically indicate a terminal node, including its
associated region and observations, atomic model, and assessment
results.

\subsection{Differential trees}

In this paper we relate the methodology of tree modeling to difference and
change detection. By following the general methodology described in
Section~\ref{sectree}, a differential tree is built from \textit{multiple}
data sets to discover distinguishing patterns between them. Its atomic
model is for observations from all data sets, but only the parameters
that account for differences in distribution are of direct interest and
examined by a homogeneity test. In particular, we will use the
Poisson distribution for the count of observations at each level of the
response in each data set to form the atomic model, and contrast the event
rates in all data sets with a likelihood ratio test of homogeneity.

This new change detection method differs from those
described in Section~\ref{secchange-detection}, in its nonparametric
nature and applicability to high-dimensional data. As for other
families of
tree models, it has the advantages of fast training (relative to most other
data mining models), easy handling of different types of variables and
dealing nicely with missing values. The resulting models are easily
comprehensible, which can be important for change detection, since it helps
suggest possible causes behind complicated phenomena.

\section{Building differential trees}
\label{secdifftree}

\subsection{A likelihood-based framework}
\label{seclikelihood-based}

We adopt the likelihood-based approach for building a differential tree and
for subsequent analysis. Using the likelihood for tree construction is not
rare in the literature, but it sometimes takes an implicit or approximate
form. For example, for classification trees the information gain splitting
criterion of \citet{quinlan1993} is equivalent to using the likelihood
ratio test, whereas the $\chi^2$ criterion [\citet{kass1980}] and the Gini
splitting criterion [\citet{breiman-etal1984}] are
approximations. \citet{su-etal2004} use the likelihood method, in place of
the least squares criterion, for building regression trees and obtain
simpler yet more accurate tree models in general. Using the likelihood
method for tree construction gives results a statistical interpretation,
deals with splitting and pruning in one framework, and permits the
handling of many\vadjust{\goodbreak}
families of atomic models in a coherent way. For building differential
trees we take one further step, by making use of the $p$-values of the
likelihood ratio tests. In general, this helps resolve several difficult
issues: (a) splitting in multiple ways; (b) adjusting in the presence of
missing values; (c) assessing patterns by their statistical significance;
and (d) adjusting for multiple hypothesis testing.

Within this framework, the likelihood ratio test or its statistic can also
be conveniently used to assess and compare models, even if there exist
nuisance parameters, as in the case of differential trees. We will make
extensive use of the fact that the log-likelihood ratio statistic $W$ is
asymptotically $\chi^2_\nu$, with degrees of freedom $\nu$ equal to the
number of free parameters for a simple hypothesis or the difference in the
number of free parameters for a composite one.

\subsection{Likelihood ratio test}
\label{seclrt}

The Poisson distribution with probability mass function
\[
f(n; \lambda) = e^{-\lambda} \frac{\lambda^n}{n!},\qquad \lambda> 0,
n = 0, 1, 2, \ldots,
\]
is widely used to model the number of occurrences
of an event over time or in space. Let $Y_i$ ($i= 1,2$) have the Poisson
distribution with rate $\lambda_i$. For testing homogeneity
\[
H_0\dvtx \lambda_1 = \lambda_2,
\]
the log-likelihood ratio statistic is given by
\[
W = 2 \{\log f(Y_1; Y_1) + \log f(Y_2; Y_2) - \log f(Y_1; \bar{Y}) -
\log
f(Y_2; \bar{Y})\},
\]
where $\bar{Y} = (Y_1 + Y_2)/2$. Under $H_0$, $W$ is asymptotically
$\chi_1^2$.

There are two parameters here, $(\lambda_1, \lambda_2)$, or, with
reparametrization, $(\lambda_1,\allowbreak \lambda_2 - \lambda_1)$. The focus is
on whether $\lambda_2 - \lambda_1 = 0$, while $\lambda_1$ is a nuisance
parameter.

\subsection{Atomic models}

Suppose there are $d$ data sets and the response variable has $c$
levels. For node $\tau$, let $D^\tau$ denote the data in its associated
subregion, and assume $Y_{ij}^\tau$ ($i = 1, \ldots, c, j = 1, \ldots,
d$), the number of\vspace*{1pt} observations of level $i$ in data set
$j$, is Poisson distributed with mean $\lambda_{ij}^\tau$ in that
subregion. The atomic model thus has $c \times d$ unknown parameters,
$\lambda_{ij}^\tau$, or, equivalently,
\[
\pmatrix{
\lambda_{11}^\tau& \lambda_{12}^\tau- \lambda_{11}^\tau& \cdots&
\lambda_{1d}^\tau - \lambda_{11}^\tau\vspace*{1pt}\cr
\lambda_{21}^\tau& \lambda_{22}^\tau- \lambda_{21}^\tau& \cdots&
\lambda_{2d}^\tau- \lambda_{21}^\tau\cr
\vdots& \vdots& \vdots& \vdots\cr
\lambda_{c1}^\tau& \lambda_{c2}^\tau- \lambda_{c1}^\tau& \cdots&
\lambda_{cd}^\tau- \lambda_{c1}^\tau}.
\]
Any nonzero difference in the matrix implies a \textit{distributional
difference} between the data sets. Of direct interest to us is whether all
the differences are exactly zero, while those in the first column are
nuisance parameters.\vadjust{\goodbreak}

We can hence perform a homogeneity test under the null hypothesis
%
\begin{equation}
\label{eqnh0}
H_0\dvtx \lambda_{i1}^\tau= \cdots= \lambda_{id}^\tau\qquad
\mbox{for } i = 1, \ldots, c.
\end{equation}
Letting $\bar{Y}_i^\tau= d^{-1} \sum_{j=1}^d Y_{ij}^\tau$
($i=1,\ldots,c$),
the log-likelihood ratio statistic becomes
%
\begin{equation}
\label{eqnWtau}
W(\tau; D^\tau) = 2 \Biggl\{ \sum_{i=1}^c \sum_{j=1}^d \log
f(Y_{ij}^\tau;
Y_{ij}^\tau) - \sum_{i=1}^c \sum_{j=1}^d \log f(Y_{ij}^\tau;
\bar{Y}_i^\tau) \Biggr\},
\end{equation}
which\vspace*{1pt} is asymptotically $\chi^2_{(d-1)c}$ under (\ref{eqnh0}). The test
provides evidence for preference between two settings of the atomic model.

As an example, consider the most significant pattern produced by the
differential tree shown later in Figure~\ref{figarson-difftree}. This
pattern covers $22$ other and $0$ suspicious fires in the first data set,
and $43$ other and $41$ suspicious fires in the second. The test statistic
value is
\begin{eqnarray*}
W & = &2 \{\log f(22; 22) + \log f(43; 43) - \log f(22; 32.5) -
\log
f(43; 32.5) \} \\
& & {} + 2 \{\log f(0; 0) + \log f(41; 41)
- \log f(0; 20.5) - \log f(41; 20.5) \} \\
& \approx&63.75,
\end{eqnarray*}
which yields a $p$-value of $1.4 \times10^{-14}$ under $\chi_2^2$.

The appropriate atomic model depends on the problem under study. By
assuming equal rates, null hypothesis (\ref{eqnh0}) implies that all data
sets were obtained under the same exposure, for example, over time
periods of
equal length. While this applies to our analysis presented below due to our
special partitioning of the data set on an annual basis, one could also
consider the case where exposures are different. If the exposures are
known, say, $e_j$ for data set $j$, one needs to modify $H_0$ to
%
\begin{equation}
\label{eqnh0p}
H_0'\dvtx
e_1 \lambda_{i1}^\tau= \cdots= e_d \lambda_{id}^\tau\qquad
\mbox{for } i = 1, \ldots, c,
\end{equation}
where $\lambda_{ij}^\tau$ is a rate per unit exposure. Reassigning
$\bar{Y}_i^\tau= \sum_{i=1}^d e_j Y_{ij}^\tau/ \sum_{i=1}^d e_j$,
we have
\[
W'(\tau; D^\tau) = 2 \Biggl\{ \sum_{i=1}^c \sum_{j=1}^d \log
f(Y_{ij}^\tau;Y_{ij}^\tau) - \sum_{i=1}^c \sum_{j=1}^d \log
f(Y_{ij}^\tau; e_j \bar{Y}_i^\tau)
\Biggr\},
\]
which is also asymptotically $\chi^2_{(d-1)c}$.

If, however, the exposures are unknown, it is impossible to test a null
hypothesis of type (\ref{eqnh0}) or (\ref{eqnh0p}). Instead, one can
investigate if every data set has the same distribution for the proportions
of all response levels, namely,
%
\begin{equation}
\label{eqnh0pp}
H_0''\dvtx p_{i1}^\tau= \cdots= p_{id}^\tau\qquad \mbox{for } i
= 1,
\ldots, c,
\end{equation}
where $p_{ij}^\tau$ is the probability an observation in data set $j$
is of
level $i$. Therefore, one can assume that $(Y_{1j}^\tau, \ldots,
Y_{cj}^\tau)^\T$ has a multinomial distribution\vadjust{\goodbreak} with probabilities
$(p_{1j}^\tau, \ldots, p_{cj}^\tau)^\T$. The log-likelihood ratio statistic
is
\[
W''(\tau; D^\tau) = 2 \Biggl\{ \sum_{i=1}^c \sum_{j=1}^d n_j^\tau
\log
\ph_{ij}^\tau- \sum_{i=1}^c \sum_{j=1}^d n_j^\tau\log\ph_i^\tau
\Biggr\},
\]
where $n_j^\tau= \sum_{i=1}^c Y_{ij}^\tau$, $\ph_{ij}^\tau=
Y_{ij}^\tau/
n_j^\tau$ and $\ph_i^\tau= \sum_{j=1}^d Y_{ij}^\tau/ \sum_{j=1}^d
n_j^\tau$. Under $H_0''$, $W''$~is asymptotically $\chi^2_{(c-1)(d-1)}$.

Throughout our study, we assume that the underlying distribution of counts
is Poisson distributed, and only the null hypothesis (\ref{eqnh0}) and the
resulting statistic (\ref{eqnWtau}) are used. In general, altering the
atomic model alters the family of differential trees being considered, but
the framework for analysis remains the same.

\subsection{Subtree models}
\label{secsubtree}

Denote by $T^\tau$ the subtree rooted at node $\tau$ and by~$\Tt
^\tau$ the
set of its terminal nodes. The log-likelihood ratio statistic for
$T^\tau$
is given by
%
\begin{equation}
\label{eqnWT}
W(T^\tau; D^\tau) = \sum_{t \in\Tt^\tau} W(t; D^t).
\end{equation}
The statistic $W(T^\tau; D^\tau)$ is approximately $\chi^2$, with degrees
of freedom given by the sum of the degrees of freedom of the individual
terms.

\subsection{Splitting}
\label{secsplitting}

We only consider univariate binary splits, which use data information most
efficiently, allow surrogate splitting in the presence of missing values,
and treat numerical variables no differently from ordinal ones. We further
turn categorical variables into ordinal ones by using their pre-given order
of levels, instead of considering all possible combinations. This avoids
the overfitting introduced by level grouping, which can be severe when
a~categorical variable has many levels. It is also helpful when the
pre-given levels are partially ordinal.

Without missing values, the primary split at node $\tau$ is determined by
%
\begin{equation}
\label{eqnprimary-split}
s^\tau_* = \argmax_{s\in S^\tau} W(T_s^\tau; D^\tau),
\end{equation}
where $T_s^\tau$ is the two-child-node tree defined by a univariate binary
split $s$ and $S^\tau$ is the set of all candidate splits at node
$\tau$. For $S^\tau$, we consider every explanatory variable and every
midpoint between two consecutive distinct values of the variable from all
data sets, but we exclude the cases where small subsets (having less
than a
total of $5c$ observations, by default) are produced. With $s^\tau_*$, all
data sets are split accordingly and the tree is grown with two new child
nodes. The splitting process starts with a single node for all data and
proceeds in a top-down, recursive style, until a stop-splitting criterion
is met, for example, too few observations left.

To find the primary split in the presence of missing values, a slight
adjustment is made using $p$-values, which takes account of different\vadjust{\goodbreak}
sample sizes caused by missing values. For the $k$th variable at node
$\tau$, denote by $n_k^\tau$ the number of observations without missing
values and by $p_{k*}^\tau$ the smallest $p$-value of all likelihood ratio
tests for the $n_k^\tau$ observations. The adjusted $p$-value is given by
%
\begin{equation}
\label{eqnpja}
\pt_{k*}^\tau= p_{k*}^\tau+ \gamma\sqrt{p_{k*}^\tau(1-p_{k*}^\tau
) /
n_k^\tau},
\end{equation}
where $\gamma> 0$ is a constant, which is defaulted to $2$ in our
implementation. Similar in spirit to the 1-SE rule of
\citet{breiman-etal1984}, the adjustment tends to favor variables with
fewer missing values.

To determine the correct branch for an observation when the primary
splitting variable at a node has a missing value, a surrogate split can be
used, as described in \citet{breiman-etal1984}, Section 5.3. A surrogate
split is made using a different variable, chosen so that the surrogate
split is as similar to the primary split as possible for the observations
without missing values. We measure the similarity of two splits by the
number of common observations in their resulting subsets. An ordered list
of surrogate splits can be constructed according to their similarities to
the primary split.

\subsection{Pruning}
\label{secpruning}

The pruning of an initially grown tree is necessary for removing spurious
subtrees. It works in a bottom-up style, by choosing either the atomic
model at an internal node or its subtree model. To do this, one could use
the cost-complexity measure, which here is just the log-likelihood
penalized by the degrees of freedom. For a subtree, it is defined as
\[
W_\alpha(T^\tau; D^\tau) = W(T^\tau; D^\tau) + \alpha\dof(\Tt
^\tau),
\]
where $\dof(\Tt^\tau)$ is the number of degrees of freedom for all the
atomic models in $\Tt^\tau$ and $\alpha$ the complexity parameter. Note
that the atomic model at a~node is just a~tree with a~single node, so its
cost-complexity measure is
\[
W_\alpha(\tau; D^\tau) = W(\tau; D^\tau) + \alpha\dof(\tau).
\]
The pruning criterion is as follows:
%
\begin{equation}
\label{eqnpruning-W}
\mbox{Choose the model with the larger value of } W_\alpha.
\end{equation}
The value of $\alpha$ can be determined by a model selection criterion,
such as AIC or BIC, or cross-validation. In principle, replacing a subtree
with its root node implies that the event frequencies cannot be
\textit{further} differentiated between the data sets in all subregions.

If the main goal for building a differential tree is to find the most
significant differences between data sets, we can simply preserve the most
significant patterns in a constructed tree. Let $\pmin(\tau)$ be the
$p$-value of the hypothesis test performed at the node $\tau$, for
example, the
likelihood ratio test based on the statistic (\ref{eqnWtau}); and
$\pmin(T^\tau)$ be the smallest $p$-value of all hypothesis\vadjust{\goodbreak} tests performed
at the atomic nodes of the subtree $T^\tau$. The new pruning criterion
is as follows:
%
\begin{equation}
\label{eqnpruning-pmin}
\mbox{Choose the model with the smaller value of } \pmin.
\end{equation}
By doing so, each subtree preserves the node with the most significant
pattern and keeps it as a terminal node. This also facilitates the
$p$-value adjustments, as described in Section~\ref{secadjust}.

In addition, one may set up a threshold $p$-value, say, $p_{\mathrm{cut}}$,
such that a~subtree is cut off directly if its $\pmin(T^\tau) \ge
p_{\mathrm{cut}}$. It helps remove the less significant patterns, while
keeping the most significant ones. The tree model may thus be greatly
simplified and can be interpreted more easily. In our implementation, we
set $p_{\mathrm{cut}} = 10^{-6}$ as default. For the arson case study, this
roughly corresponds to $p'' = 0.25$; see Section~\ref{secadjust} for the
definition of $p''$.

\subsection{Pseudo code}

To serve as a summary, Algorithm~\ref{algdifftree} gives the pseudo code
of the recursive function that we implemented for building a differential
tree from two data sets.

\begin{algorithm}[!htb]
\caption{Differential Tree Construction (from Two Data
Sets)} \label{algdifftree}
\textbf{function} \difftree($D1$, $D2$)
\begin{algorithmic}[1]
\REQUIRE Data sets $D1$ and $D2$
\STATE Create a terminal node $\tau$
\STATE Compute $W(\tau; (D1, D2))$ and $p$-value, using (\ref{eqnWtau})
\IF{too few observations in $D1$ and $D2$}
\return$\tau$
\ENDIF
\STATE Label $\tau$ as an internal node
\FOR{each predictor variable}
\STATE Find all potential splits from its distinct values in $D1$ and
$D2$
\STATE Compute $W(T_s^\tau; (D1, D2))$ for each potential split $s$,
using (\ref{eqnWT})
\ENDFOR
\STATE Find the primary split $s^\tau_*$, using (\ref{eqnprimary-split})
[or (\ref{eqnpja}) in the presence of missing values] \STATE Find all
surrogate splits of $s^\tau_*$
\STATE Use $s^\tau_*$ (and possibly
surrogate splits) to partition $D1$ into $(D1_{\mathrm{left}},
D1_{\mathrm{right}})$ and $D2$ into $(D2_{\mathrm{left}},
D2_{\mathrm{right}})$
\STATE$\tau$\$left $=$
\difftree($D1_{\mathrm{left}}$, $D2_{\mathrm{left}}$) \STATE
$\tau$\$right $=$ \difftree($D1_{\mathrm{right}}$, $D2_{\mathrm{right}}$)
\IF{$\tau$ is preferred over $T^\tau$ by (\ref{eqnpruning-pmin})}
\STATE
Discard $\tau$\$left and $\tau$\$right and label $\tau$ as a terminal
node
\ENDIF
\return$\tau$
\end{algorithmic}
\end{algorithm}

\section{A primary study of the arson case}
\label{secarson}

\subsection{Setup}
\label{secsetup}

In this section we compare two approaches to solving the arson
problem. Both use \texttt{label} as the response; one utilizes traditional
classification trees, and the other builds a differential tree
directly. For our
case study the latter is more efficient at discovering differential
patterns.

We divide the arson data set into two subsets, covering two time periods,
2004--2005 and 2006--2007, respectively (and reset 1/Jan/2006 to
\texttt{day}${}={}$1 and similarly the days after). The two subsets contain,
respectively, $318$ and $386$ fire incidents, of which $80$ and $91$ are
suspicious. Both suspicious and other fires are included in the study,
because a maliciously-set fire is not necessarily labeled suspicious or
vice versa, and because it illustrates the application of the method to a
multiple category problem. In Section~\ref{secfurther-studies} we apply
the proposed method to detect changes in the frequencies of suspicious
fires only, and of fire incidents with a different categorization.

\subsection{Two classification trees}
\label{secarson-dt}

To discover differential patterns, let us first consider building two
classification trees, one from each subset, and then testing all the
patterns induced from a classification tree against the other subset. The
rationale is that classification trees, if constructed properly, are
consistent estimators of the underlying distributions
[\citet{breiman-etal1984}, Chapter 12] and thus their differences are also
consistent for
estimating the true distributional differences. Note that each individual
classification tree is only built to model the underlying relation between
the response and explanatory variables for a \textit{single} data set and
thus inevitably may include patterns that are common with the other,
for example,
for seasonal effects.

\begin{figure}

\includegraphics{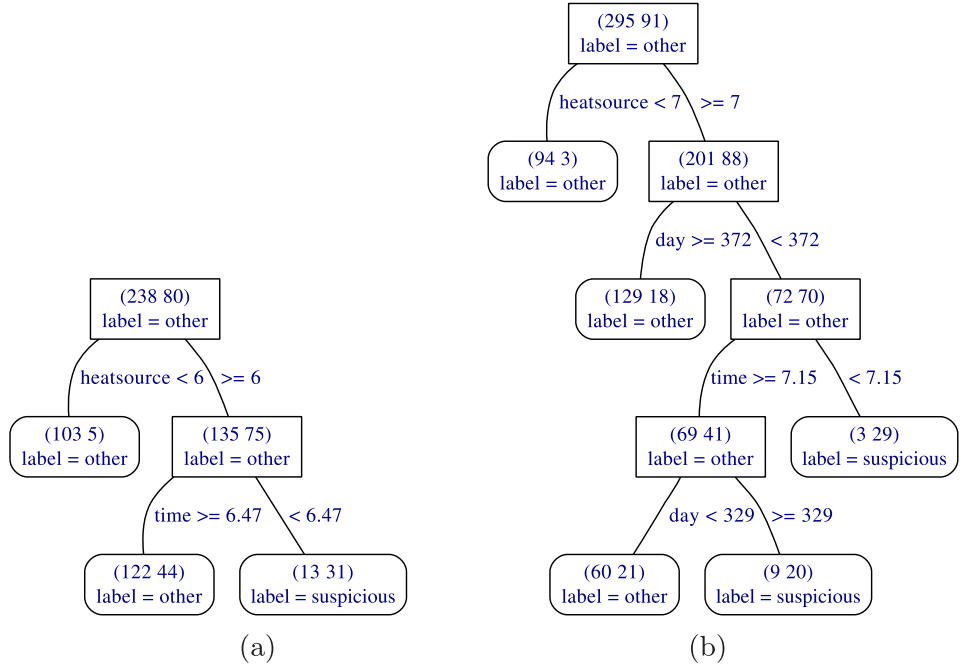}

\caption{Classification trees built from fire incidents in Blenheim during:
\textup{(a)} 1/Jan/2004--\allowbreak31/Dec/2005 and \textup{(b)} 1/Jan/2006--31/Dec/2007.
Inside the parentheses at a node are the numbers of
observations for each response level, here ``other'' and
``suspicious.''}
\label{figarson-tree}
\end{figure}

\begin{table}
\tabcolsep=0pt
\caption{Patterns obtained from each of the two classification trees and
tested by their covered observations in both subsets}
\label{tabarson}
\begin{tabular*}{\tablewidth}{@{\extracolsep{\fill}}lccccccc@{}}
\hline
& \multicolumn{3}{c}{\textbf{2004--2005}}
& \multicolumn{3}{c}{\textbf{2006--2007}} & \\[-4pt]
& \multicolumn{3}{c}{\hrulefill} & \multicolumn{3}{c}{\hrulefill}\\
\textbf{Pattern} & \textbf{\texttt{Other}} & \textbf{\texttt{Suspicious}}
& \textbf{Proportion} & \textbf{\texttt{Other}}
& \textbf{\texttt{Suspicious}} & \textbf{Proportion}
& \multicolumn{1}{c@{}}{$\bolds{p}$\textbf{-value}} \\
\hline
(a)
& \multicolumn{3}{c}{\textit{Training set}}
& \multicolumn{3}{c}{\textit{Test set}} & \\
[4pt]
1 & 103 & \hphantom{0}5 & 0.046 & \hphantom{0}90 & \hphantom{0}2 & 0.022 & $3.3 \times10^{-1}$ \\
2 & 122 & 44 & 0.265 & 183 & 55 & 0.231 & $1.2 \times10^{-3}$\\
3 & \hphantom{0}13 & 31 & 0.705 & \hphantom{0}22 & 34 & 0.607 & $2.9 \times10^{-1}$\\
[4pt]
(b) & \multicolumn{3}{c}{\textit{Test set}} &
\multicolumn{3}{c}{\textit{Training set}} & \\
[4pt]
4 & 105 & \hphantom{0}7 & 0.062 & \hphantom{0}94 & \hphantom{0}3 & 0.031 & $3.2 \times10^{-1}$\\
5 & \hphantom{0}77 & 38 & 0.330 & 129 & 18 & 0.122 & $3.4 \times10^{-5}$\\
6 & \hphantom{0}36 & 17 & 0.321 & \hphantom{0}60 & 21 & 0.259 & $3.9 \times10^{-2}$ \\
7 & \hphantom{0}15 & \hphantom{0}0 & 0.000 & \hphantom{00}9 & 20 & 0.690 & $4.5 \times10^{-7}$\\
8 & \hphantom{00}5 & 18 & 0.783 & \hphantom{00}3 & 29 & 0.906 & $2.1 \times10^{-1}$\\
\hline
\end{tabular*} \vspace*{24pt}
\end{table}

\begin{figure}

\includegraphics{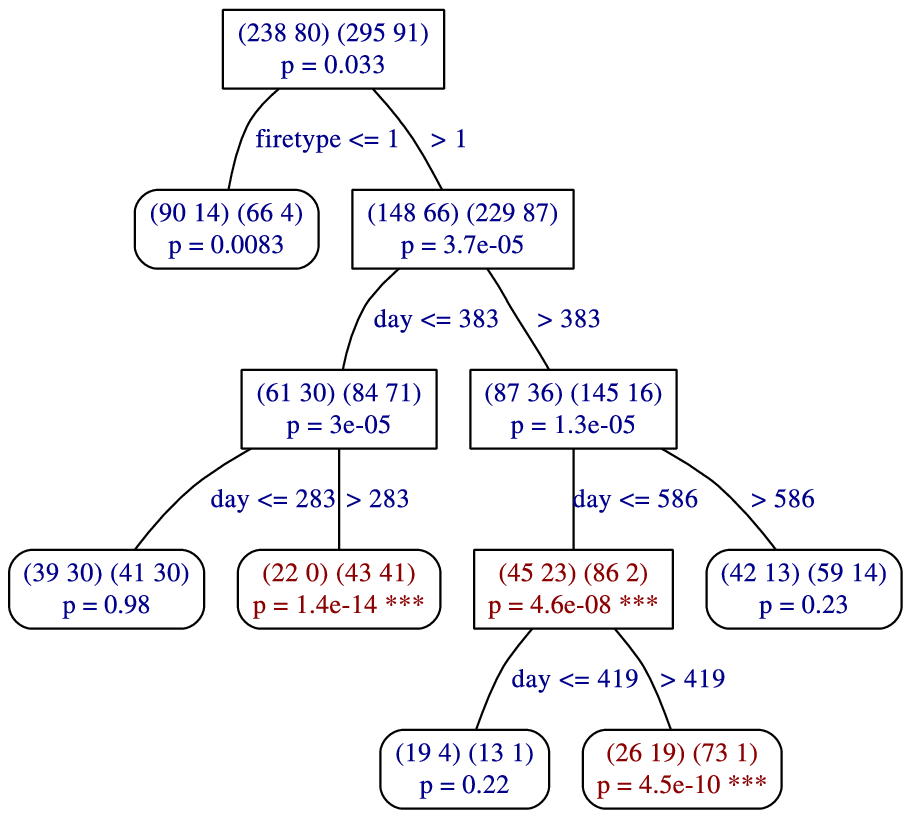}

\caption{Differential tree built directly by contrasting the fire
incidents from 1/Jan/2004--\allowbreak31/Dec/2005 with those from
1/Jan/2006--31/Dec/2007. Each pair of parentheses at a~node contains the numbers of
observations for all response levels in a data set, and~for the arson
data here (\texttt{\#other}, \texttt{\#suspicious}). Any $p$-value less
than $10^{-5}$ is marked ``***.''}
\label{figarson-difftree}
\end{figure}

We use the R package ``\texttt{rpart}'' [\citet{therneau-atkinson1997}] for
classification tree construction. The classification tree built from the
first subset is shown in Figure~\ref{figarson-tree}(a). The tree identifies
three situations or patterns, as also listed in the upper part of
Table~\ref{tabarson}, in ascending order of their estimated
proportions of
suspicious fires. To eliminate the patterns that are irrelevant to
differences, we test them against the second subset, using
(\ref{eqnWtau}). Hence, the remaining significant patterns can only be
attributed to the distributional differences between the two subsets. After
this screening, only pattern 2 remains significant, with a $p$-value of
$1.2 \times10^{-3}$. Nonetheless, its significance is mainly due to an
increase of ``other'' in 2006--2007, rather than a change in the frequency
of suspicious fires.

Analogously, we can find patterns from the second subset and test them
against the first subset. The classification tree built from the second
subset is shown in Figure~\ref{figarson-tree}(b). It contains five patterns,
as listed in the lower part of Table~\ref{tabarson}. Pattern~$7$ is the
most significant, with a $p$-value of $4.5 \times10^{-7}$, which
corresponds to a remarkable increase of $20$ suspicious fires and appears
to be related to the arson case. Specifically, it indicates a significant
increase in the proportion\vadjust{\goodbreak} of suspicious fires between day 329
(25/Nov/2006) and day 371 (6/Jan/2007), for time after 7:12~am, with heat
source that includes cigarettes/matches/candles. It possesses a very
different characteristic from pattern 8, which classifies fires as highly
suspicious that occur between 0:00 am and 7:12 am, due to heat
source${}\ge{}$7. However, with a $p$-value of $0.21$, pattern 8 does not suggest a
change, although it merits further investigation by itself. Pattern 5 is
also highly significant but corresponds to a decrease of $38- 18 =20$
suspicious fires, as well as a substantial increase of $129 - 77 = 52$
other fires; this change occurred after day 371 (6/Jan/2007).

\subsection{One differential tree}
\label{secarson-difftree}

\begin{figure}

\includegraphics{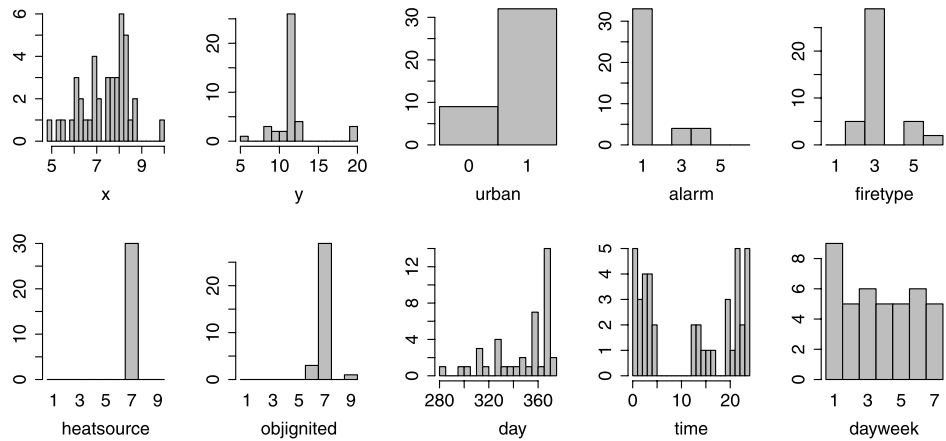}

\caption{Histograms/barplots for the $41$ suspicious fires covered by the
most significant pattern.}\vspace*{-2pt}
\label{figarson-histall2}
\end{figure}

A differential tree between the two subsets is constructed, as shown in
Figure~\ref{figarson-difftree}, which contains six terminal nodes. The most
significant, with a $p$-value of $1.4 \times10^{-14}$, appears to relate
directly to the arson case. Specifically, it suggests that a change has
occurred between day 284 (11/Oct/2006) and day 383 (18/Jan/2007), with all
types but property fires. The change is due to a substantial increase of
$41 - 0 = 41$ suspicious fires, as well as an increase of $43-22=21$ other
fires. To gain more information about these $41$ suspicious fires, the
histograms/barplots for all predictor variables are shown in
Figure~\ref{figarson-histall2}. These fires are exclusively due to
\texttt{heatsource}${}={}$7 ($=$cigarettes/matches/candles), mainly of
\texttt{firetype}${}={}$3 ($=$Vegetation), largely distributed along a
horizontal strip (variable \texttt{y}), and having an increasing trend over
time (variable \texttt{day}).

The second most significant pattern has a $p$-value of $4.5 \times10^{-10}$
and specifies a situation where there is a decrease in the number of
suspicious fires and yet an increase in the number of other fires. This
change took place between day 420 (24/Feb/2007) and day 586 (9/Aug/2007).

Note that the general conclusions drawn here are similar to those in
Section~\ref{secarson-dt}. This is not really surprising, since both
methods provide consistent estimators for detecting differences between the
two underlying distributions. However, we should also notice that the
patterns found by the differential tree, that searches for changes directly
by ignoring irrelevant patterns, are statistically much more significant
[even using the properly adjusted $p$-value (\ref{eqnperm-arson}) or
(\ref{eqnperm-arson2})]. This suggests that differential trees are the
more efficient approach to change detection. There must therefore be
situations where real changes can be detected by the differential tree
approach, but not by the other, and especially so when a data set contains
many significant patterns not attributable to changes.\vspace*{-1pt}

\section{Performance assessment and enhancement}\vspace*{-1pt}
\label{secperform}

\subsection{Significance adjustment}
\label{secadjust}

Since ``significant patterns'' can always be found with an exhaustive
search, one should consider their possible spuriousness. In the following,
we consider adjusting $p$-values using the Bonferroni and the permutation
method.

The Bonferroni method is the simplest and most conservative. For $m$ tests
performed, it adjusts their smallest $p$-value, say, $p$, by
%
\begin{equation}
\label{eqnbonf}
p' = \min\{m p, 1\}.\vadjust{\goodbreak}
\end{equation}
For building the differential tree shown in Figure~\ref{figarson-difftree},
there are 13,414 tests performed in total. This includes all candidate
splits examined at the splitting stage, including those at the nodes that
are cut off later, but not any comparisons at the pruning stage due to their
irrelevance in determining the minimum $p$-value. Its adjusted
$p$-value for
the most significant pattern is thus
%
\begin{equation}
\label{eqnbonf-arson}
p' = 13\mbox{,}414 \times1.4 \times10^{-14} \approx1.9 \times10^{-10},
\end{equation}
which remains highly significant, despite the conservativeness of the
method.

The permutation method adjusts a $p$-value by using it as a statistic and
is based on the fact that, under the null hypothesis, the adjusted
$p$-value has the uniform distribution on $[0,1]$. The $p$-value to be
adjusted can be either~$p$ or $p'$ in (\ref{eqnbonf}), and, for the arson
data, it does not appear to make much difference. In general, we are
inclined to use $p'$ since it guards against the situation where an
extremely small $p$-value is produced through an exhaustive search. The
empirical null distribution can be obtained by permuting either the entire
data under investigation, which may nonetheless contain irregular changes
and hence reduce the power of detection, or, better, some comparable,
``clean'' historical data. For the arson case, we choose to permute the
entire data here, and later in Section~\ref{secsequential} some historical
data.

Specifically, our adjustment proceeds as follows. Each observation in the
two subsets created in Section~\ref{secsetup} is randomly reallocated to
either the first or second biennial period by tossing a fair coin (without
changing its date within a biennial period), thus ensuring the null hypothesis
(\ref{eqnh0}) is satisfied. This shuffling destroys all distributional
differences between the two periods, but preserves all the relations among
the variables such as geographical clusters and seasonal effects. For each
pair of random subsets, a differential tree is constructed, and a minimum
$p$-value obtained and adjusted by (\ref{eqnbonf}). With $R$ ($=$1000
throughout the paper) random replications, $R$ copies of the $p'$-value are
obtained and ordered into $p'_{(1)} \le\cdots\le p'_{(R)}$, whose
self-adjusted values are, respectively, $1/(R+1), \ldots, R/(R+1)$, namely,
their expectations under the null hypothesis. Letting $p'_{(0)} = 0$ and
$p'_{(R+1)} = 1$, a new $p'$ can be adjusted by interpolation:
%
\begin{equation}
\label{eqnperm}
p'' = \frac{j + r}{R+1}\qquad \mbox{if } p'_{(j)} \le p' \le
p'_{(j+1)},\qquad j = 0, \ldots, R,
\end{equation}
where $r = (p' - p'_{(j)}) / (p'_{(j+1)} - p'_{(j)})$.

From\vspace*{1pt} the $1000$ differential trees constructed, we obtained $p'_{(1)} = 8.4
\times10^{-6}$, and, therefore, the permutation adjusted $p$-value for the
most significant pattern in the differential tree shown in
Figure~\ref{figarson-difftree} is
%
\begin{equation}
\label{eqnperm-arson}
p'' = \frac{1.9 \times10^{-10} / 8.4 \times10^{-6}}{1001} \approx2.3
\times10^{-8}.
\end{equation}
This is still an extremely small $p$-value, indicating that it is highly
unlikely that this discovered pattern occurred purely by chance.

\begin{figure}

\includegraphics{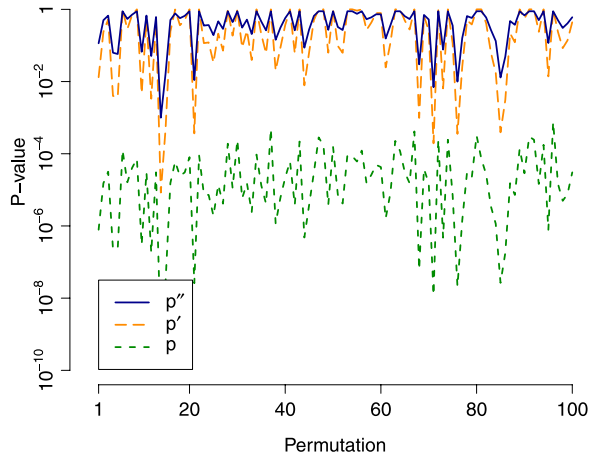}

\caption{Minimum $p$-values and adjustments in the differential trees
constructed from the first $100$, out of $1000$, random permutations of
the $4$-year data.}
\label{figperm}\vspace*{-3pt}
\end{figure}

Figure~\ref{figperm} shows the minimum $p$-values from the first $100$
permutations, along with their adjustments. Despite the pure randomness of
the permutations, the minimum $p$-values produced by differential trees are
remarkably small, indicating the necessity of adjustment. What
surprises us
most, as can also be seen in results given later, is that $p''$ is almost
always larger than~$p'$, because the Bonferroni adjustment is theoretically
the most conservative. How could this happen? We think that the reason may
lie in the fact that, conditional on the data, patterns with the smallest
$p$-values are sought in a~deterministic manner, and this has some
similarity to a deterministic optimization process, which violates the
underlying assumption of \textit{randomness} for multiple hypothesis
testing. If this is true, it has profound implications for many modern
statistical methods of modeling and hypothesis testing that involve
extensive data manipulation to find the ``best'' solutions. The bias
introduced by such data manipulation may be very high, so high that even
the most conservative method can fail to bound it.\vspace*{-3pt}


\subsection{Bootstrap aggregating}
\label{secbagging}

One problem with tree models is instability [\citet{breiman1996c}], which
means that a small perturbation in the data may result in a tree with a
substantially different structure. In general, an unstable estimator tends
to exhibit high variation and low predictive power. For differential trees,
this is relevant for discovered differential patterns and their
significance levels. Instability, however, can be reduced, often
considerably, by using meta-learning techniques, such as boosting
[\citet{freund-schapire1997}], bagging (bootstrap aggregating)
[\citet{breiman1996b}] or random forests\vadjust{\goodbreak} [\citet{breiman2001}], which resort
to building a number of models by perturbing the data. In the following we
consider the bagging technique to stabilize the estimation of the minimum
$p$-value in a differential tree.

To use bagging on the two subsets described in Section~\ref{secsetup}, we
draw a bootstrap sample from each subset and build a differential tree
from the pair of bootstrap samples, which gives a minimum $p$-value and its
Bonferroni adjustment $p'$. This is repeated $B$ ($=$50 throughout the
paper) times. The median of the $B$ resulting $p'$-values is then taken
as the
bagging estimate of the Bonferroni-adjusted minimum $p$-value. From a random
run, we obtained an estimate $p' = 1.8 \times10^{-11}$.

To find the empirical null distribution of the bagging estimator for
permutation adjustment, $1000$ random replications of the $4$-year data
were produced with random allocations to the two biennial periods, in a
similar fashion to Section~\ref{secadjust}. The above bagging estimator
is then applied to each replication. The five-number summary of the
resulting $p'$-values is $(6.4 \times10^{-7}, 2.7 \times10^{-5}, 5.5
\times10^{-5}, 1.2 \times10^{-4}, 5.9 \times10^{-4})$. Thus, the
permutation adjusted $p$-value is
%
\begin{equation}
\label{eqnperm-arson2}
p'' = \frac{1.8 \times10^{-11} / 6.4 \times10^{-7}}{1001} \approx
2.8 \times10^{-8}.
\end{equation}
As we shall see in Section~\ref{secsim}, the bagging-based adjusted
$p$-values are less variable and, when there exist true differences, tend
to be smaller than those that are produced without using bagging.

One problem with bagging is that it does not produce one but many trees,
which loses the interpretability of a single differential tree. A possible
remedy is to associate a $p$-value with each observation, for example,
using the
median $p''$-value of all patterns that cover the observation. Then we know
which observations are associated with changes, and how
significantly. Areas containing observations with small $p$-values can
perhaps be derived subsequently.

\subsection{A simulation study}
\label{secsim}

In order to gauge the efficiency and stability of the differential tree
method, we conducted a simulation study and made use of the arson data
in a
way that mimicked the arson case. To produce random data for two biennial
periods, all $318$ ($238$ other and $80$ suspicious) fire incidents in
2004--2005 are duplicated once and then randomly reallocated to either the
first or second biennial period by coin tossing (as in
Section~\ref{secadjust}). We did not include the data in 2006--2007 to
avoid contamination. Then we added $n_\Delta\in\{0,10,\ldots,50\}$
distinctive fire incidents to the second biennial period, randomly drawn
from those 2006--2007 incidents covered by the most significant pattern
discovered in Section~\ref{secarson-difftree}, in the proportions of
$30\%$ other and $70\%$ suspicious fires. For each $n_\Delta\in
\{0,10,\ldots,50\}$, $100$ such data sets were generated, and thus $100$
(without bagging) and $100 \times50$ (with bagging) differential trees
were built. To adjust $p$-values, $1000$ permutations were carried out both\vadjust{\goodbreak}
with and without bagging, thus producing $1000 + 1000 \times50$
differential trees. A total of 81,600 differential trees were built in
the simulation study.

\begin{figure}

\includegraphics{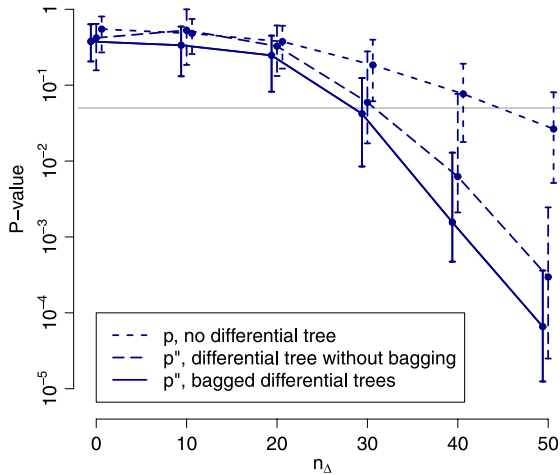}

\caption{Each vertical line segment represents the central $50\%$ interval
of an empirical distribution obtained from $100$ $p$- or $p''$-values, and
a solid point the median. Some line segments are slightly shifted
horizontally for distinguishing purposes. The horizontal line is where
$p$-value${}={}$0.05.}
\label{figsim}\vspace*{-3pt}
\end{figure}

Figure~\ref{figsim} shows summaries of the $p$-values ($p$ or $p''$)
produced by three methods: a direct evaluation using (\ref{eqnWtau})
without building any differential tree (which gives the same $p$-value as
the root node of a differential tree), building one differential tree, and
building differential trees with bagging. The central $50\%$ interval of
the empirical distribution of the $p$-value is plotted for each case. It
can be seen that, when $n_\Delta= 0$, all three $p$-values appear to
conform well with the uniform distribution on $[0,1]$. We can use the
medians of these $p$-values to gauge efficiency and the widths of the
central $50\%$ intervals to gauge stability. As $n_\Delta$ increases, each
$p$-value decreases, and at an increasing rate. The directly evaluated $p$,
however, decreases slowly and this approach, on average, is unable to
detect the change at the $5\%$ significance level until $n_\Delta
\approx
44$. An intuitive explanation is that a dramatic change deep under the
surface may only manifest as ripples on the surface, that is, at the root
node. Also, multiple changes may even cancel out the effects of one
another and leave no trace on the surface, as is the case of the
differential tree shown in Figure~\ref{figarson-difftree}. By contrast,
with differential trees it ``dives'' down and seeks differences between the
data sets in increasingly smaller areas. The method can thus uncover local
differences more efficiently and, for the arson problem, is able to start
detecting the change for $n_\Delta\approx31$ (without bagging) and
$\approx$27 (with bagging). As $n_\Delta$ increases, there are also
clearly widening gaps between the $p$-values produced by the direct
evaluation method and the differential tree methods. For $n_\Delta= 50$,
the median $p$ is only $2.6 \times10^{-2}$,\vadjust{\goodbreak} being barely significant,
while the median $p''$ is $3.0 \times10^{-4}$ (without bagging) or $6.6
\times10^{-5}$ (with bagging). It is also clear that the bagging technique
helped reduce instability and increase efficiency. The arson case has
$n_{\Delta}$ close to $60$, which we could not include in the simulation
study since it requires $42$ suspicious fires but the most significant
pattern has only $41$. However, with a~visual extrapolation of the curves
in Figure~\ref{figsim} to where $n_\Delta= 60$, it should be clear
that the
proposed method is quite effective for discovering the changes in the arson
case.\vspace*{-3pt}

\section{Sequential detection}
\label{secsequential}

The method developed above can also be used in a sequential detection
manner. Let us consider comparing the data of the two consecutive annual
periods immediately before a ``detection'' day (the first day after the
two year period). With the quadrennial data
available, we start the detection from 1/Jan/2006, by building a
differential tree that compares the two time periods,
1/Jan/2004--31/Dec/2004 and 1/Jan/2005--31/Dec/2005, and build new differential trees
by shifting the detection day at intervals of seven days, until all data
have been examined. From every tree constructed the smallest $p$-value
is extracted and
adjusted by the Bonferroni and permutation methods, using (\ref{eqnbonf})
and (\ref{eqnperm}). The empirical null distribution of the minimum
$p$-value in a differential tree that is needed by the permutation adjustment
is obtained by permuting $1000$ times the historical fire incidents that
occurred during 1/Jan/2004--31/Dec/2005. We have also produced an
empirical null distribution by permuting random halves of all the
quadrennial data and found that the resulting adjusted $p$-values are only
slightly larger, due to the contamination of the irregular changes in the
latter two years. The conclusions, however, remain largely the same. To use
bagging, one only needs to replace each single differential tree described
above with $50$ trees obtained under bootstrap sampling
(Section~\ref{secbagging}).

\begin{figure}

\includegraphics{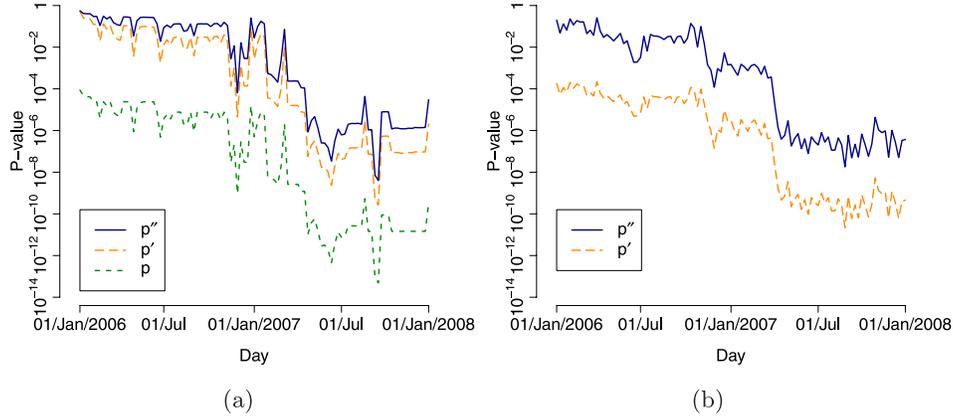}

\caption{Minimum $p$-values and their adjustments in sequential detection:
\textup{(a)} without bagging; \textup{(b)} with bagging. In particular, $p$ is the
minimum $p$-value in a differential tree, $p'$ the Bonferroni adjustment
of $p$, and $p''$ the permutation adjustment of $p'$.}
\label{figpvalues}
\end{figure}

The results are shown in Figure~\ref{figpvalues}. The sequential detection
results with bagging shown in Figure~\ref{figpvalues}(b) are clearly more
stable than those without bagging in Figure~\ref{figpvalues}(a). From
Figure~\ref{figpvalues}(a), after the initial $45$ weeks with
basically no
significant change discovered and a smallest $p''$-value of $0.025$, a
sudden decrease of the $p''$-value occurred on detection day $316$
(12/Nov/2006) with $p'' = 0.0040$. This is clearly a sign that some
significant change(s) have occurred in the underlying data-generating
mechanism. Similar conclusions can be drawn from the more stable estimates
in Figure~\ref{figpvalues}(b).

By monitoring the change of (adjusted) $p$-values, it is
straightforward for
an online system to set up different levels of warning in an easily
comprehensible sense.

\section{Using different responses}
\label{secfurther-studies}

\subsection{Using only suspicious fires}

\begin{figure}[b]

\includegraphics{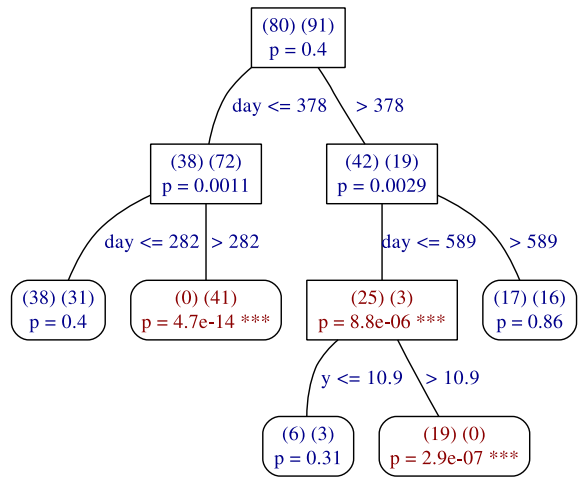}

\caption{Differential tree built from using suspicious fires only.}
\label{figarson-difftree2}
\end{figure}

Instead of using both suspicious and other fires as done in the study so
far, we can use suspicious fires only. Figure~\ref{figarson-difftree2}
displays the differential tree,\vadjust{\goodbreak} built analogously to that in
Figure~\ref{figarson-difftree}, from the two biennial subsets.
Interestingly, the two most significant patterns are comparable in both
trees: one concerning a substantial increase of suspicious fires during
almost the same time period and the other a decrease of suspicious fires
after it. Note that one cannot use the classification tree approach here,
since the response variable has only one level.

\begin{figure}

\includegraphics{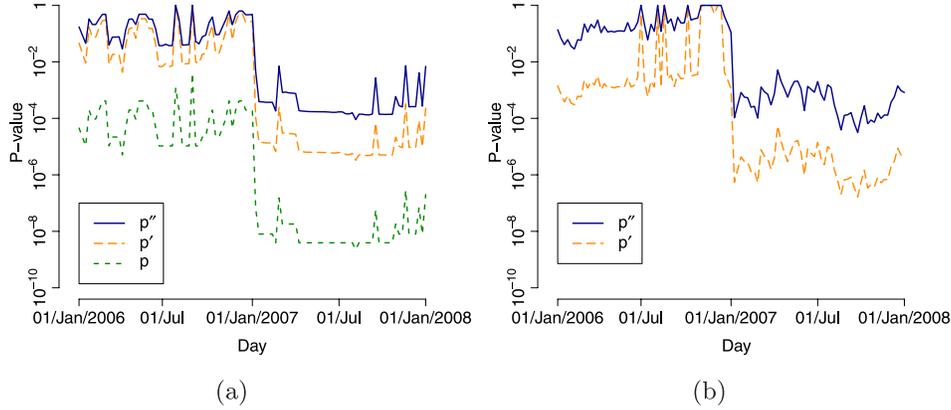}

\caption{$p$-values in sequential detection using suspicious fires only:
\textup{(a)} without bagging; \textup{(b)} with bagging.}
\label{figsusp-pvalues}
\end{figure}

\begin{figure}[b]

\includegraphics{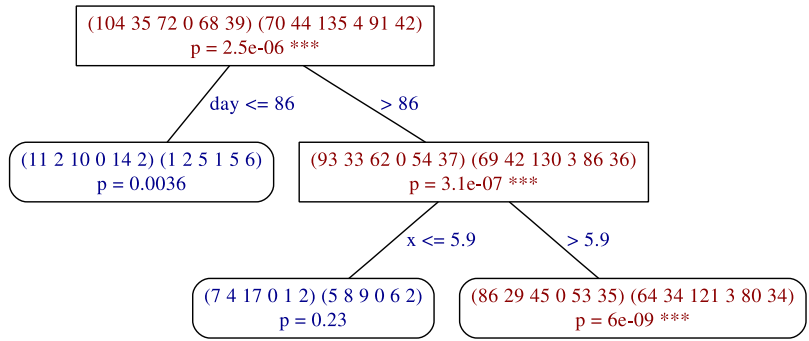}

\caption{Differential tree built with a different response variable.}
\label{figarson-difftree3}
\end{figure}

The minimum $p$-values and their adjustments for sequential detection are
plotted in Figure~\ref{figsusp-pvalues}. The sudden change has also been
successfully detected, but at a delayed date, as compared with that in
Figure~\ref{figpvalues}. This is because most suspicious fires
occurred in
the latter part of the biennial period (see the histogram of
\texttt{day} in
Figure~\ref{figarson-histall2}), and because in the earlier part of
the time
period there is an increase of fire incidents that are not labeled
``suspicious,'' which are thus excluded from the study here. In this case,
including all fire incidents is preferable---it gives an earlier warning!

\subsection{Using an alternative response variable}

One can also use a different response variable, as if for a
general surveillance, in total ignorance of what has happened. Let us
this time treat the variable \texttt{firetype} as the response. The
differential tree built from the two biennial subsets is shown in
Figure~\ref{figarson-difftree3}. This tree appears to be less informative
and its most significant pattern is also less significant, as compared with
the trees shown in Figures~\ref{figarson-difftree} and
\ref{figarson-difftree2}. However, this discovered pattern is still
remarkably significant, showing that the difference is mainly due to an
increase of vegetation fires, jumping from $45$ cases to $121$ for
\texttt{day}${}>{}$86 and $\mbox{\texttt{x}}> 5.9$.

\begin{figure}

\includegraphics{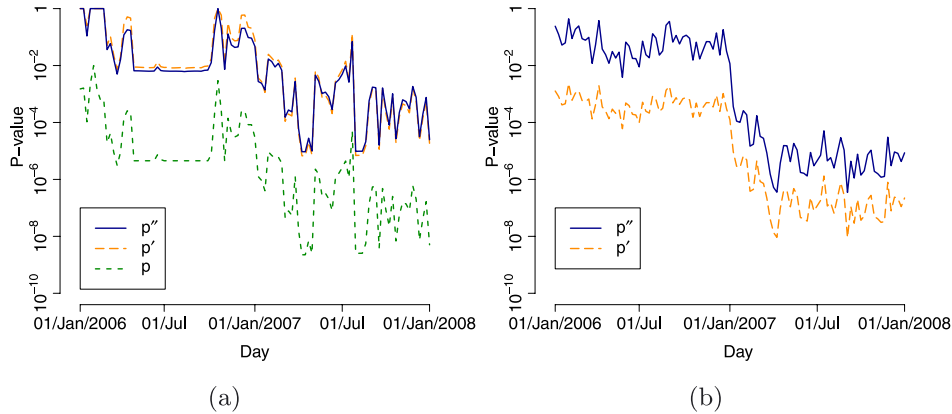}

\caption{$p$-values in a sequential detection using \texttt{firetype} as
response: \textup{(a)} without bagging; \textup{(b)} with bagging.}
\label{figalt-pvalues}
\end{figure}

The minimum $p$-values and their adjustments obtained via sequential
detection are shown in Figure~\ref{figalt-pvalues}. It is clear that the
change has also been detected, at a later date and less dramatically than
that in Section~\ref{secsequential}.

These results are perhaps the most one could hope for when
conducting a general surveillance.

\section{Concluding remarks}
\label{secsummary}

There are two main new ideas in our proposed method for change or
difference detection. One is to contrast data sets and model their
distributional differences and the other to use tree models to uncover
local, irregular changes and provide interpretable results. We followed the
general methodology for tree construction. Variants with improved
performance likely exist, as in the literature for other families of
the tree
model. Extensions to other types of difference detection seem fairly
straightforward.

Building a differential tree is reasonably fast. With our
implementation in
R \mbox{[\citet{R2011}]}, it took, respectively, $5.0$, $1.4$ and $6.8$
seconds to
build the trees shown in Figures~\ref{figarson-difftree},
\ref{figarson-difftree2} and~\ref{figarson-difftree3}, on a workstation
with a 2.93~GHz CPU. This made possible the demanding numerical studies
reported earlier. If implemented in FORTRAN or C, it is likely much faster.

Finally, we give a rationale for using differential trees in a complex
environment. A general alternative is to compare the data with a reference
model that can be either exactly known, which is virtually impossible
in a
complex environment, or estimated from a reference data set, just as we did
in Section~\ref{secarson-dt}. Since building a model from one data set and
testing it against the other can waste data information on discovering
patterns irrelevant to differences and we are essentially comparing two
data sets, why do not we just build one model that directly describes their
differences? This is exactly what a differential tree does.

\section*{Acknowledgments}

We are grateful to the Editor, Associate Editor and two reviewers for their
helpful and constructive comments that led to many improvements in the
manuscript.

\begin{supplement}
\sname{Supplement A}
\stitle{Data}
\slink[doi]{10.1214/12-AOAS548SUPPA} 
\slink[url]{http://lib.stat.cmu.edu/aoas/548/arson.csv}
\sdatatype{.csv}
\sdescription{The file \texttt{arson.csv} contains the Arson data that is
described in Section~\ref{secintro} and used in the analysis throughout the paper.
The meanings of the variables and the values they
take on are available in Table~\ref{tabvariables}.}
\end{supplement}

\begin{supplement}
\sname{Supplement B}
\stitle{Software}
\slink[doi]{10.1214/12-AOAS548SUPPB}
\slink[url]{http://lib.stat.cmu.edu/aoas/548/difftree.R}
\sdatatype{.R}
\sdescription{The file \texttt{difftree.R} contains the R code for carrying
out the analysis in the paper.}
\end{supplement}

%

\printaddresses

\end{document}